\documentclass{amsart}

\usepackage{amssymb}
\usepackage{amscd}

\begin{document}

\title{On the Clausius theorem}

\author{Alexey~Gavrilov}
\email{gavrilov@lapasrv.sscc.ru}

\address{\newline Institute of Computational Mathematics and Mathematical
Geophysics;\newline ~Russia,~Novosibirsk}

\begin{abstract}
We show that for a metastable system there exists
a theoretical possibility of a violation of the Clausius inequality without a
violation of the second law.
Possibilities of experimental detection of this hypothetical violation
are pointed out.
\end{abstract}

\maketitle

The question which will be considered here is the possibility of a violation
of the Clausius inequality
$$\oint\frac{\delta Q}{T}\le 0.\eqno{(1)}$$
We do not discuss here the second law of thermodynamics. It remains
beyond any doubt. The aim is to
watch closely the proof of (1) which is not as rigorous as it seems.
As we'll see, this proof is based on some assumptions which theoretically
may be wrong for a metastable system. Our consideration avoids
statistical mechanics, it is mostly thermodynamic, in other words,
phenomenological. 

\section{Clausius theorem}

The well-known Clausius theorem states that the inequality (1) is valid
for any closed system undergoing a cycle.
The proof may be found in some textbooks on thermodynamics
\footnote{Forgive the author: he was unable to find a book where
the proof have been written thoroughly. There was the gabbles or the references
to inaccessible (for him) sources.}.
Consider a system connected to a heat reservoir at a constant absolute
temperature of $T_0$ through a reversible cyclic device (e.g., Carnot engine)
(fig. 1).

\begin{picture}(350,120)
\put(170,0){{\bf fig.1}}
\put(20,60){heat}
\put(12,50){reservoir}
\put(205,80){system}
\put(10,15){\framebox(40,100)}
\put(300,15){\framebox(40,100)}
\put(200,60){\framebox(40,40)}
\put(310,60){work}
\put(302,50){reservoir}

\put(125,80){\circle{40}}
\put(50,80){\line(10,0){55}}
\put(145,80){\line(10,0){55}}
\put(70,84){$Q_0$}
\put(165,84){$Q$}
\put(270,84){$W$}
\put(270,44){$W^{\prime}$}
\thicklines
\put(240,80){\line(10,0){60}}
\put(125,60){\line(0,-10){20}}
\put(125,40){\line(10,0){175}}

\end{picture}

Then $$\oint\frac{\delta Q}{T}=\frac{Q_0}{T_0},$$
where $Q_0$ is the amount of heat taken from the heat reservoir.
The combined system (=system+reversible device) operates in a cycle,
hence the possibility of $Q_0>0$ is forbidden by the Kelvin-Plank
formulation of the second law.

This reasonong is implicitely based on the following assumption which may
be called a postulate: {\it there is no interaction between two closed
systems without heat or work exchange}. If it's so, we can always replace
the interaction with surroundings by the interaction with heat and work
reservoirs and no problems arise.

Now, let us suppose that the postulate is wrong. Consider two interacting
systems both of which undergo a cyclic process. Applying the usual scheme
(fig. 7 below), we can prove the inequality
$$\oint\frac{\delta Q_1}{T_1}+\oint\frac{\delta Q_2}{T_2}\le 0.$$
But to prove (1) for one of the systems we have to prevent their interaction.
We can replace the heat exchange between the systems by the heat
exchange between each of the systems and a heat reservoir. We can replace
the work exchange by the interaction with a work reservoir. But if the
systems still interact somehow then we can do nothing. We can't make
one of them undergo the same cycle without interaction with another,
so we can't prove (1) for any of them. As we'll see, the theoretical
possibility of exotic (not obeying the postulate) interactions
appears for metastable systems. Thus, we can't exclude the possibility
of the Clausius inequality violation.

In the next three sections the models are considered which break the inequality
(1). Are these models purely imaginary or have something to do with reality
is a question to which the author can't give a proper answer.
He supposes that the inequality may be broken actually.
But it is impossible to prove it with thermodynamic arguments.
This calls for stronger means, such as an experiment, or, maybe,
kinetic theory. In section 5 we will describe this hypothetical
violation in thermodynamic terms.
In the last two sections two of the models (including the Szilard engine)
will be considered in more details.

\section{Xenium engine}

Let xenium be an imaginary gas whose
molecules may be in two different states on the same energy level.
Assume further that spontaneous transitions between these states
are very rare, but two sufficiently close molecules
may {\it exchange} their states. It doesn't matter whether a gas with
these properties exists. The question is whether we can {\it imagine} it
without breaking the principal laws of nature. The author sees no obstacles.

The xenium in a closed volume is a metastable system which may be
considered as a mixture of two components, say, xenium-1 and xenium-2
(in the same way as ortho- and parahydrogen). If two volumes with xenium-1
in one and with xenium-2 in another are separated by such a thin membrane that
molecules may interact through it, then the gas in each volume will be
mixed with another kind of xenium. This process is very similar to the usual
diffusion through the membrane, but both of volumes remain closed systems and
the number of molecules in each volume remains.

Now consider a cylinder provided with a piston and closed with a thin
membrane on the other side. In the middle of the cylinder there is
a membrane of another kind (selective) . It is thick but
porous and it is permeable for xenium-1 but not for xenium-2.
This device is permanently attached to a heat reservoir. There are also two
large reservoirs with xenium-1 in one and with xenium-2 in another to
either of which the cylinder can be attached or not, as needed (fig. 2).

\begin{picture}(360,150)
\put(170,0){{\bf fig.2}}
\put(300,15){\line(0,10){100}}
\put(310,15){\line(0,10){100}}

\put(317,90){x}
\put(317,80){e}
\put(317,70){n}
\put(317,60){i}
\put(317,50){u}
\put(317,40){m}
\put(317,25){2}

\put(225,60){xenium-1}
\put(220,50){+ xenium-2}
\put(120,60){xenium-1}

\put(300,115){\line(-1,1){10}}
\put(310,115){\line(-2,1){20}}

\put(290,125){\line(-10,0){70}}
\put(225,127){{\small thin membranes}}

\put(175,115){\line(-1,1){10}}
\put(165,125){\line(-10,0){70}}
\put(100,127){{\small selective membrane}}

\thicklines
\put(310,115){\line(10,0){20}}
\put(310,15){\line(10,0){20}}
\put(175,15){\line(0,10){100}}
\put(100,18){\line(0,10){94}}
\put(90,18){\line(0,10){42}}
\put(90,112){\line(0,-10){42}}
\put(90,112){\line(10,0){10}}
\put(90,18){\line(10,0){10}}
\put(90,60){\line(-10,0){70}}
\put(90,70){\line(-10,0){70}}
\put(50,15){\line(10,0){250}}
\put(50,115){\line(10,0){250}}
\end{picture}

At the beginning the piston is pushed to the selective membrane and the space
between two membranes is filled with a mixture of xenium-1 and xenium-2
in equal proportion. Then the cylinder is connected to the xenium-1 reservoir
for some time, so the gas bacames enriched with this fraction. In the next step
the piston moves out (isothermal expansion). After that the device is switched
from the xenium-1 reservoir to the xenium-2 one, until the numbers of molecules
in two states {\it in all the cylinder} became equal. Then xenium
reservoir is removed and the piston is pushed to the initial position
(isothermal compression).

Note that the device (without  heat and xenium reservoirs) is a closed system
undergoing a cycle. The expansion takes place at higher (partial) pressure
of the working gas (xenium-1) than the compression. Hence a positive amount of
work is produced at the expence of heat taken from heat reservoir.
So, this device takes a positive amount of heat despite the inequality (1).
Obviously, no contradiction to the second law arises. The standard proof
doesn't work here because of the exotic interaction between the device
and the xenium reservoirs.

\section{Szilard engine}

The Szilard engine is an imaginary device proposed by L.Szilard in 1929 [1].
It consists of a cylinder with two pistons and a removable partition.
The working gas consists of a single molecule (fig. 3).

\begin{picture}(350,120)
\put(170,0){{\bf fig.3}}
\put(60,20){\line(10,0){220}}
\put(60,90){\line(10,0){100}}
\put(280,90){\line(-10,0){100}}
\put(162,50){\framebox(16,69)}
\put(80,22){\line(0,10){66}}
\put(260,22){\line(0,10){66}}
\put(80,22){\line(-10,0){10}}
\put(260,22){\line(10,0){10}}
\put(80,88){\line(-10,0){10}}
\put(260,88){\line(10,0){10}}
\put(70,22){\line(0,10){28}}
\put(270,22){\line(0,10){28}}
\put(70,88){\line(0,-10){28}}
\put(270,88){\line(0,-10){28}}
\put(270,60){\line(10,0){40}}
\put(70,50){\line(-10,0){40}}
\put(70,60){\line(-10,0){40}}
\put(270,50){\line(10,0){40}}

\put(105,43){\circle{17}}

\end{picture}

This device, permanently connected to a thermostat, is provided with
Maxwell's demon, which is an automatic control system operating as following.
At the beginning, the demon places the partition in the middle of the cylinder,
dividing it into two equal chambers. It takes a look at the molecule, finding out in which
half of the cylinder it has been trapped, and places a piston in the other half.
This doesn't involve any work, since the piston is pushed against nothing.
After that, the demon removes the partition and the piston moves back.
In this process (isothermic expansion) the one-molecule working gas
produses work, which amounts to $k_BT\log 2.$ Repeating the cycle, the engine
converts heat to work seemingly as the perpetuum mobile of the second kind.

The explanation which becames now standard was proposed by C.H.Bennett
(see, e.g., [2]). The demon receives one bit of information per cycle.
To complete the cycle it has to return the memory to the initial state,
i.e. to erase this information. According to the Landauer's principle,
the erasure of information must be accompanied by a heat generation
amounts to $k_B T\log 2$ per bit. So, all the produced work will be spended.

This short description avoids many details. The reader may find more
about the Szilard engine and the Landauer's principle in [2,3,4].
Despite some criticism, the author is sure
that the Bennett's explanation is perfectly correct. But he is not going to
prove it. He wants only to point out that {\it if} it is correct
{\it then} the Szilard engine violates the Clausius inequality
(not the second law!). We can consider the "mechanical part" of this device,
i.e. the engine without the demon, as a closed system. This system
{\it returns} to the original state after the cycle (while the demon may not).
In this cycle it receives a positive amount of heat despite (1).
Again, the standard proof fails because of the exotic interaction between
the engine and the demon.

\section{Smoluchowski pump}

Now we consider another kind of device controlled by a Maxwell's demon.
It is a pipe with two removable partitions, filled with a gas (fig. 4).

\begin{picture}(350,120)
\put(170,0){{\bf fig.4}}
\put(10,20){\line(10,0){330}}
\put(10,80){\line(10,0){90}}
\put(120,80){\line(10,0){100}}
\put(340,80){\line(-10,0){100}}
\put(102,22){\framebox(16,61)}
\put(222,62){\framebox(16,57)}

\put(61,37){\circle{17}}
\put(183,54){\circle{17}}

\end{picture}

The pressure of a gas is assumed to be small such that the mean number
of molecules in the volume between partitions is about unity (say, $10^{-1}$).
The demon is watching this volume to see whether there are any molecules
inside. While the volume is empty it keeps the left partition removed and the
right one inserted. When a molecule comes into the volume it places the
left partition into the pipe and removes the right one. Note that molecules
may enter the volume by the left partition and come out by the right one only.
Hence this device, which will be called a Smoluchowski pump works as a
Smoluchowski trapdoor, moving the gas from left to right.

It is well- known that the Smoluchowski trapdoor fails to work because of
the heat fluctuations. However, there exists a possibility that the pump
works. The difference is the "intelligent" nature of the demon
which, as we'll see later, means the metastability of the demon as
a thermodynamic system. From Bennett's point of view, the demon receives
information, hence the Smoluchowski pump's demon may work as well as
the Szilard engine's one, until it's memory is full.

It's easy to see that this pump is forbidden by the Clausius inequality.
Take two vessels filled with a gas, connected with two pipes with a
Smoluchowski pump in the middle of one and with a turbine in the middle
of another (fig. 5). When the pump works it makes the pressure in the
right vessel greater than the pressure in the left one. Then the gas
flows through the turbine delivering work. Unlike a usual pump
the Smoluchowski one performs no work on the gas, hence all the
produced work is at the expence of heat taken from the envirenment,
which may be a thermostat. Obviously, the inequality (1) is broken.
However, no contradiction to the second law arises for the same reasons as
for a Szilard engine.

\begin{picture}(350,140)
\put(170,0){{\bf fig.5}}
\put(20,20){\framebox(50,100)}
\put(280,20){\framebox(50,100)}
\put(150,95){\framebox(60,20)}
\put(180,40){\circle{40}}
\put(180,40){\circle{15}}
\put(70,102){\line(10,0){80}}
\put(70,108){\line(10,0){80}}
\put(210,108){\line(10,0){70}}
\put(210,102){\line(10,0){70}}
\put(70,60){\line(10,0){110}}
\put(70,54){\line(10,0){95}}
\put(280,37){\line(-10,0){80}}
\put(280,43){\line(-10,0){80}}
\put(170,120){pump}
\put(170,62){turbine}
\put(230,110){$\rightarrow$}
\put(230,45){$\leftarrow$}

\end{picture}

The Smoluchowski pump seems as unrealistic as the Szilard engine.
However, the recent progress in the nanoelectronics gives us a hope to build
a Smoluchowski electron pump, operating with electrons instead of molecules.
Consider a device similar to one investigated in [5] (fig. 6). It consists
of two metallic leads connected by tunnel barriers to two quantum dots
(pay attention to the backward bias voltage). The left dot may be closed
or opened by tuning a voltage on a gate. As in [5], the device is provided
with a sensor checking in which dot the electron appears (assuming
the both dots may not be occupied).

\begin{picture}(350,140)
\put(170,0){{\bf fig.6}}
\put(0,60){\line(10,0){140}}
\put(0,90){\line(10,0){140}}
\put(350,90){\line(-10,0){120}}
\put(350,60){\line(-10,0){120}}
\put(140,60){\line(0,10){30}}
\put(230,60){\line(0,10){30}}
\put(150,60){\framebox(30,30)}
\put(190,60){\framebox(30,30)}
\put(70,30){\line(10,0){110}}
\put(300,30){\line(-10,0){115}}
\put(70,30){\line(0,10){30}}
\put(300,30){\line(0,10){30}}
\put(180,20){\line(0,10){20}}
\put(185,10){\line(0,10){40}}
\put(150,100){\line(10,0){30}}
\put(165,100){\line(0,10){40}}
\put(60,70){source}
\put(270,70){drain}
\put(170,120){gate}

\end{picture}

If the bias voltage $V$ is comparable with the value
$k_BT/e$ ($\approx 30 mV$ at the room temperature) then electrons may
move forward and backward due to the charge fluctuations. The control device
(Maxwell's demon) closes the left dot whenever an electron occures
in the right one. So, electrons move only in one direction and the
device works as a pump.
To break the Clausius inequality
is is enough to achieve the inequality $W<eV,$ where $W$ is the work done on
the electron. Theoretically, $W$ may be measured directly, but in practice
it may be a problem.
The author is not competent to answer whether this pump may be built presently.
We should leave this question to the experts.

\section{Entropy of a metastable system}

The reader may regard the models described above as purely imaginary.
The main goal was to show the {\it logical } groundlessness of the
Clausius theorem. However, it doesn't mean too much.
A statement may be true under much weaker
assumptions that those under which it has been proved.
There is a lot of examples. To show the real possibility of the violation of
the Clausius inequality the author needs stronger arguments than those
 he can make presently.
Nevertheless, we may consider the following question. The models are not
logically inconsistent, so we may {\it suppose} they are real.
How can we describe them in thermodynamic terms? The key property is
the {\it metastability}.

First of all, we have to define the reversible process precisely.
An isolated system undergoes a reversible process if this one can proceed in
reverse direction. A closed system undergoes a reversible process if
it is or may be a part of an isolated system under the same condition.
This definition, in slightly different terms, may be found in many
sources. (Some authors add {\it extra} condition: the system
should be at equiliblum. In author's view, it is superfluous.
Take a diamond, put it into a thermostat and wait for a year.
Does it undergo a reversible process or not?)
Now, applying this notion of a reversible process to a metastable system
with equiliblum surroundings, we
can define it's entropy as usual: $dS=\frac{\delta Q_{rev}}{T}$
(as we'll see later, non-equilibrum surroundings should be handled with caution).

Phenomenologically,
what is the difference between the equiliblum state and the metastable one?
The equilibrum state depends solely on the temperature $T$ and the set
of extensive parameters (such as the volume, the mole number etc.).
Then, the internal energy $U$ and the entropy $S$ are the functions of these variables.
Excluding the temperature, we can write $U=U(S,\dots,X_k,\dots),$
where $X_1,\dots,X_k,\dots$ are extensive parameters.
The metastable state may as well depend on some additional parameter(s).
It is called the {\it order parameter}, denoting by $\eta$.
In this case, $U=U(S,\dots,X_k,\dots,\eta).$

Fix all the parameters except for the entropy (or, equivalently, temperature).
Then the internal energy change is the result of the heat exchange only,
which is a reversible process.
We may suppose the heat source (drain) to be a thermostat, which is a
system at equilibrum. Then, by the definition, $dS=\frac{dU}{T},$
or $(\frac{\partial U}{\partial S})_{X,\eta}=T.$ Thus, in a reversible process,
$$dU=TdS+\delta W+\delta M,$$
where $\delta W=\sum_{k}\frac{\partial U}{\partial X_k}dX_k$ is, obviously,
the work and $\delta M=\frac{\partial U}{\partial \eta}d\eta$
is the term which is usually called a mass action. Note that this term
does not depend on the particular choice of the order parameter.

The case $\delta M\neq 0$ is {\it very} unusual for a closed system.
However, we may not exclude this term {\it a priory}. It is convenient to
introduce the value $\delta S^{\prime}=-\frac{\delta M}{T},$
which will be called a heatless entropy.
Now, in an arbitrary (not necessary reversible) process,
$dU=\delta Q+\delta W\le TdS+\delta W+\delta M,$  hence
$$dS\ge\frac{\delta Q}{T}+\delta S^{\prime}.\eqno{(2)}$$
This is a corrected version of the (differential) Clausius inequality.
As usual, inequality becames an equality for a reversible process.
From (2) it is clear that $\delta S^{\prime}$ means the
amount of entropy taken from the surroundings without heat.

Consider two closed metastable systems interacting with each other in a
reversible process. We can connect both of them to a heat reservoir and
a work reservoir such that the composed system becames isolated (fig 7).
If we denote by $dS_0$ the heat reservoir entropy differential,
then we obtain
$$dS_1=\frac{\delta Q_1}{T_1}+\delta S_1^{\prime},\,
dS_2=\frac{\delta Q_2}{T_2}+\delta S_2^{\prime},
dS_0=-\frac{\delta Q_1}{T_1}-\frac{\delta Q_2}{T_2}.$$
But $dS_1+dS_2+dS_0=0,$ hence $\delta S_1^{\prime}+\delta S_2^{\prime}=0.$
We came to a conclusion that heatless entropy does not appear or disappear
but may only be transferred from one system to another.
An important consequence is that to make heatless entropy transfer possible
the both of the interacting systems should be metastable.

\begin{picture}(350,170)
\put(170,0){{\bf fig.7}}
\put(20,20){\framebox(310,20)}
\put(20,140){\framebox(310,20)}
\put(80,60){\framebox(50,30)}
\put(130,75){\line(10,0){80}}
\put(145,77){interaction}

\put(210,60){\framebox(50,30)}
\put(105,115){\circle{30}}
\put(235,115){\circle{30}}
\put(60,40){\line(0,10){75}}
\put(280,40){\line(0,10){75}}
\put(60,115){\line(10,0){30}}
\put(280,115){\line(-10,0){30}}
\put(105,40){\line(0,10){20}}
\put(235,40){\line(0,10){20}}
\put(105,90){\line(0,10){10}}
\put(235,90){\line(0,10){10}}
\put(105,130){\line(0,10){10}}
\put(235,130){\line(0,10){10}}
\put(130,30){work reservoir}
\put(130,150){heat reservoir}
\put(90,70){system 1}
\put(220,70){system 2}

\end{picture}

Integrating (2), we have for a cyclic process
$$\oint\frac{\delta Q}{T}+S^{\prime}\le 0.\eqno{(3)}$$
When $S^{\prime}<0,$ the inequality (1) is not necessary.
For an isothermal cyclic process $dU-TdS$ is a differential.
Then, integrating the inequality $\delta W\ge dU-TdS+T\delta S^{\prime},$
we have in this case
$$W\ge TS^{\prime}.\eqno{(4)}$$
Thus, an engine may work with a single heat source if it gives heatless
entropy somewhere.

\section{Xenium engine's work}

Now we can apply our theory to the xenium engine. We consider xenium as a
two-component gas. For the total internal energy in a reversible
process we have the equation
$$dU=TdS-PdV-P_2dV_2+\mu dN+\mu_2 dN_2,$$
where $P,P_2,V,V_2,\mu,\mu_2,N$ and $N_2$ are the partial pressure,
the volume, the chemical potential and the mole number of xenium-1 and xenium-2
respectively. This equation may be simplified, because the volume $V_2$
doesn't change and $N+N_2=const$:
$$dU=TdS-PdV+\Delta\mu dN,\, \Delta\mu=\mu-\mu_2.$$
Obviously, $\delta W=-PdV,\,\delta M=\Delta\mu dN,$
hence $\delta S^{\prime}=-\frac{\Delta\mu}{T}dN.$ This term describes
heatless entropy transfer between the engine and the xenium reservoir.
Easy computation gives for a cyclic process
$W=TS^{\prime}$, as it should be under the reversibility assumption.
Thus, the engine produces work and gives heatless entropy to
xenium reservoirs.

There is a minor remark. The equality
$\delta S^{\prime}=-\frac{\Delta\mu}{T}dN$ has been obtained under the
reversibility assumption. Dropping this assumption, we may take only
the inequality $\delta S^{\prime}\le-\frac{\Delta\mu}{T}dN.$
For usual two-component metastable system, such as low- temperature
hydrogen, we have the same inequality but it becames meaningless
because really $\delta S^{\prime}=0.$

\section{Szilard engine's work}

The work produced by a Szilard engine may be found by the formula (4),
as for a xenium engine. However, in this case the question how the
entropy is transferred from the engine (to the demon) becames more
complicated.

First of all, there should be two metastable systems. The first one is a
Maxwell's demon. It is worthy to be remembered that {\it any} system with
memory is metastable almost by the definition. The equilibrum memory
would be a subject to thermodynamic fluctuations, in which case
no one could use it.
In view of statistical mechanics, when the partition is inserted
into the cylinder, the one-molecule gas becames a non-ergodic system: it's phase
space is divided in two parts. Thus, this gas should be considered as a
metastable one.

At the beginning, the demon's memory is supposed to be empty and it's entropy
$S_d=0.$ The partition is removed and entropy of a gas is equal to
$S_g=k_B\log(2V)+S_0,$ where $V$ is the half-volume of the cylinder and
$S_0$ depends on the temperature only. For the sake of simplicity we set
$S_0=0.$ The process may be divided into three steps.

{\bf Step 1} The partition is inserted. It doesn't change entropy but the gas
becames (potentially) metastable.
The possibility of heatless entropy transfer appears.

{\bf Step 2} The demon findes out in which half of the cylinder
the molecule is closed. The states of  gas and demon becames statistically
dependent: memory state depends on the  molecule position.
The correct approach to this case is to consider the total entropy $S_{g+d},$
which is no more the sum of $S_g$ and $S_d.$ 
The measurement is a reversible process [3], hence the total entropy doesn't
change: $S_{g+d}=k_B\log(2V).$

{\bf Step 3}  The demon pushes the piston and removes the partition. This is
a reversible process again. The states of gas and demon became statistically
independent again. The gas is equilibrum (no more metastable) and
it's volume is $V.$ Hence, $S_g=k_B\log(V).$ The demon's memory is in one
of two random states, and $S_d=k_B\log 2.$ As a result, heatless entropy amounts
to $S^{\prime}=k_B\log 2$ has been transferred from the engine to the demon.

According to (4), the Szilard engine may produce work amounts to
$k_B T\log 2$ per cycle. But according to the same inequality
the demon has to convert the produced work (and perhaps some more)
into heat to operate in a cycle. Thus, this device may not be a perpetuum
mobile.

\end{document}